\documentclass[aps,showpacs,prc,nofootinbib,floatfix]{revtex4}

\usepackage{graphicx}

\usepackage{latexsym}
\usepackage{hhline}
\usepackage{amssymb}
\usepackage{wasysym}
\usepackage{epsfig}

\sloppypar

\newcommand{\be}{\begin{equation}}
\newcommand{\ee}{\end{equation}}
\newcommand{\bea}{\begin{eqnarray}}
\newcommand{\eea}{\end{eqnarray}}
\newcommand{\ba}{\begin{array}}
\newcommand{\ea}{\end{array}}
\newcommand{\bi}{\begin{itemize}}
\newcommand{\ei}{\end{itemize}}

\newcommand{\ffh}{\frac{5}{2}}

\begin{document}

\title{A coupled-channel analysis of $K\Lambda$ production\\
in the nucleon resonance region.
\footnote{Supported by Forschungszentrum Juelich}}

\author{V. Shklyar\footnote{On leave from Far Eastern State University,
                       690600 Vladivostok, Russia}}
\email{shklyar@theo.physik.uni-giessen.de}
\author{H. Lenske}
\author{U. Mosel}
\affiliation{Institut f\"ur Theoretische Physik, Universit\"at Giessen, D-35392
Giessen, Germany}

\begin{abstract}
A unitary coupled-channel effective Lagrangian model is applied to 
the combined analysis of the $(\pi,\gamma) N \to K\Lambda$  
reactions in the  energy region up to 2 GeV. To constrain the resonance 
couplings to the  $K\Lambda$ final state the recent photoproduction 
data obtained by the SAPHIR, SPring-8, and CLAS groups are included 
into the calculations. The main resonance contributions
to the process stem  from the $S_{11}(1650)$, $P_{13}(1720)$, 
and $P_{13}(1900)$ states. The second bump  at 1.9 GeV seen in 
the photoproduction cross section data  is described as a coherent
sum of the resonance and background contributions. The prediction for 
the beam  polarization observable is presented.
\end{abstract}


\maketitle

\section{Introduction}
The associated strangeness  production  provides a very interesting tool
for  investigations of the nucleon resonance spectrum.
While previous experimental studies of the $\pi N \to K \Lambda$
reactions were hampered by poor statistics, recently the interest
in associated strangeness production has been rekindled by the new
photoproduction data from SAPHIR, CLAS, and SPring-8. One of the
motivations of those studies was a search for the 'missing'
resonances which might be weakly  coupled  to the $\pi N$ final
state \cite{Capstick:1998uh} and therefore are not seen in $\pi N$
scattering.

The assumption that such  'hidden' resonances can be excited in
photon-induced reactions led to the experimental study of
$K\Lambda$ and $K\Sigma$ photoproduction with the high resolution
SAPHIR spectrometer at Bonn. The first results published in 1998
\cite{Tran:1998} revealed  a  resonance-like structure in the
total $\gamma N \to K\Lambda$ cross section at 1.9 GeV. This
behaviour was explained by Penner and Mosel
\cite{Penner:2002b} as an interference pattern between the nucleon
and $t$-channel background contributions whereas Mart and Bennhold
\cite{Mart:1999} identified it with a resonance contribution from
the 'missing' $D_{13}(1960)$ state. In  a recent coupled-channel
study of the associated strangeness production
\cite{JuliaDiaz:2005} a contribution from the third $S_{11}$
resonance state is found to be necessary to describe the  CLAS
data \cite{McNabb:2003}.

A large number of other models have been developed to describe the
$K\Lambda$ data and extract the resonance couplings to these
channels. Most models  are based on the single channel formulation
of the scattering problem. They mainly differ in their treatment
of the background contributions  and number of the resonances
included \cite{Mart:1999,Mart:2004,Janssen:2001,Ireland:2004}. On
the other hand, coupled-channel models
\cite{shklyar:2004a,Penner:2002a,Penner:2002b,Chiang:2001,JuliaDiaz:2005,Lutz:2001,Usov:2005}
have been developed to simultaneously describe the pion- and
photon-induced reactions. These approaches are of advantage since
the threshold and rescattering effects in the intermediate
channels are also taken into account. The importance of a
coupled-channel description of the $K\Lambda$ photoproduction has
been demonstrated in \cite{Chiang:2001}. It has been shown  that
the contribution of the intermediate $\pi N$ channel to the total
$\gamma N \to K\Lambda$ cross section can  account for up to 20\%.

Despite of the extensive studies of the $K\Lambda$ photoproduction
the situation is far from satisfactory. Almost all models
demonstrate a good agreement with the experimental data but
predict  different resonance contributions to the process.  The
problems in the interpretation of the associated strangeness
photoproduction data in the nucleon  resonance region are well
documented \cite{Saghai:2001,Janssen:2001,Ireland:2004}.
Therefore, the central question of the resonance contribution to
the $K\Lambda$ channel is still open. Keeping that in mind, we
have performed a new study of the pion- and photon-induced
reactions within the unitary coupled-channel effective Lagrangian
approach developed in
\cite{shklyar:2004a,Penner:2002a,Penner:2002b,Penner:PhD}.

Our results for the non-strange channels and resonance parameters
extracted are presented in  \cite{shklyar:2004b}. In this paper we
continue the discussion started in \cite{shklyar:2004b} focusing
on the results on  the $K\Lambda$ production in the pion- and
photon- nucleon scattering. First, as compared to our previous
calculations \cite{Penner:2002a,Penner:2002b} the model space  has
been extended to include the contributions from the spin-$\ffh$
resonances \cite{shklyar:2004a,shklyar:2004b}. While the effect of
these states on the $\pi N \to K\Lambda$ reactions is found to be
small their contributions to the photoproduction  might be
enhanced. Secondly, new experimental data from the CLAS
\cite{McNabb:2003}, SPring-8 \cite{Zegers:2003}, and SAPHIR
\cite{Glander:2003}  collaborations have become available. This
raises the question of whether these data can be described by
already known mechanisms
\cite{Penner:2002a,Penner:2002b,shklyar:2004b} or require further
investigations  of the $K\Lambda$ reaction mechanism.

Since the polarization observables are found to be extremely
useful to distinguish between various model assumptions on the
$\gamma N\to K\Lambda$ reaction \cite{Ireland:2004},  we predict
polarization observables  which can be measured at the present
experimental facilities. This study becomes especially interesting
in the  prospect of future high resolution  data from CLAS and
SPring-8.

We start in Section \ref{compar} with a short review of the
progress made in studying $K\Lambda$ photoproduction. The main
ingredients of the applied model are discussed in Section
\ref{model}. The database and details of the calculations are
presented in Section \ref{details}.  In Section \ref{results} we
discuss the obtained results and finish with a summary.

\section{\label{compar}Overview of  the $K\Lambda$
photoproduction} Extensive studies of the $K\Lambda$
photoproduction are made in
\cite{Janssen:2001,Janssen:2001pe,Janssen:2003zv,Ireland:2004}
utilizing a tree-level description of the transition amplitude. In
their latest work \cite{Ireland:2004} Ireland, Janssen,  and
Ryckebusch attempted to distinguish between different resonance
contributions to the $K\Lambda$ channel by using a generic
algorithm analysis. Based on this procedure, these authors
conclude that a $P_{11}(1900)$ state is the most favorable
candidate for the resonant contribution to the  $K\Lambda$
photoproduction apparently seen at 1.9 GeV.  The contribution from
the $S_{11}$ and $D_{13}$ states is only weakly supported but,
nevertheless, cannot be excluded in these calculations.

Guided by the results of quark model predictions of Capstic and
Roberts \cite{Capstick:1998uh}, Mart, Sulaksono, and Bennhold  performed an
analysis \cite{Mart:2004} of the recent photoproduction data from
the SAPHIR collaboration \cite{Glander:2003}. This approach uses a
Breit-Wigner parametrization of the resonance amplitudes and,
therefore, is similar to that of
\cite{Janssen:2001,Janssen:2001pe,Janssen:2003zv,Ireland:2004}.
The  calculations of \cite{Mart:2004}  suggest a large number of
new (hidden) resonances which can contribute to the process.

The main shortcoming of the single channel Breit-Wigner models
(BW) is that the rescattering effects are missed in such
parametrizations. If one assumes a resonance which couples
strongly to the $K\Lambda$ final state, the corresponding part of
its width should be accounted for by rescattering in the
intermediate $K\Lambda$ channel. Therefore, the BW parametrization
of resonance contributions brings an ambiguity into  the calculations. This
point has been explicitly demonstrated in the work of Usov and
Scholten \cite{Usov:2005} with the example of the $S_{11}(1650)$
resonance. In the tree-level approximation to the $K\Lambda$
photoproduction amplitude the contribution from this resonance is
proportional to the product of two coupling constants $g_{\gamma
NN^*}g_{K\Lambda N^*}$. On the contrary, as soon as rescattering
effects are taken into account the final result is found to be
extremely sensitive to the value of $g_{K\Lambda N^*}$ alone, even
if the common strength $g_{\gamma NN^*}g_{K\Lambda N^*}$ is kept
constant. Therefore, the conclusions on the resonance couplings to
the $K\Lambda$ final state are different from those from  the BW
calculations.

In other words, unitarity should be maintained in any calculations
aimed to extract information on the resonance contributions to the
$K\Lambda$ channel  from experiment. Recently, models which
preserve  unitarity have  been applied to the analysis of the
$K\Lambda$ production in the pion and photon-induced reactions
\cite{Feuster:1998a,Feuster:1998b,Penner:2002a,Penner:2002b,Usov:2005,Lutz:2001,Chiang:2004,JuliaDiaz:2005}.
In the effective Lagrangian approach of Lutz, Wolf, and Friman
\cite{Lutz:2001} point-like vertices are used to describe
meson-nucleon interactions, enforcing
both unitary and analyticity. However, the assumption made about the $S$ 
and $D$-wave dominance of the reaction mechanism limits these calculations 
to energies close to the reaction threshold.

Another interesting coupled-channel approach satisfying unitarity
and analiticity is developed in
\cite{Chiang:2004,JuliaDiaz:2005,Chiang:2001}. In this model the
rescattering effects in the intermediate $\pi N$ channel are taken
into account. This is achieved by using  $\pi N$ amplitudes from
the SAID group analysis \cite{Arndt:2003}. In
\cite{JuliaDiaz:2005} the authors find a strong need for a
$S_{11}(1900)$ resonance contribution to the $K\Lambda$
photoproduction to describe the CLAS data at 1.9 GeV. Concluding
on the importance of the $\pi N$ rescattering process the authors,
however, do not check whether  other inelastic channels are
affecting the results of their calculations.

Recently, Usov and Scholten \cite{Usov:2005} presented a
coupled-channel model for the pion- and photon-induced reactions
with $\pi N$, $\eta N$, $K\Lambda$, and $K\Sigma$ in the final
state. This approach is based on the $K$-matrix formalism and
thereby is similar to the Giessen model
\cite{Feuster:1998a,Feuster:1998b,Penner:PhD,Penner:2002a,Penner:2002b,shklyar:2004a,shklyar:2004b}
 to be discussed below.
To constrain the resonance contributions to the $K\Lambda$ final
state, also $\pi N$ elastic and photoproduction data were
described with a satisfactory agreement. The main result of
\cite{Usov:2005} is that the $K\Lambda$ photoproduction above  1.7
GeV is strongly influenced by background contributions. As we will
see later, our present calculations in general support the
conclusions drawn in \cite{Usov:2005}.

\section{\label{model}Giessen Model}
The details of the model, interaction Lagrangians, and results for the non-strange channels
can be found in \cite{Penner:2002a,Penner:2002b} and \cite{shklyar:2004b} respectively.
Here, we only briefly outline the main ingredients of our model.

\begin{figure}
  \begin{center}
{\includegraphics*[width=14cm]{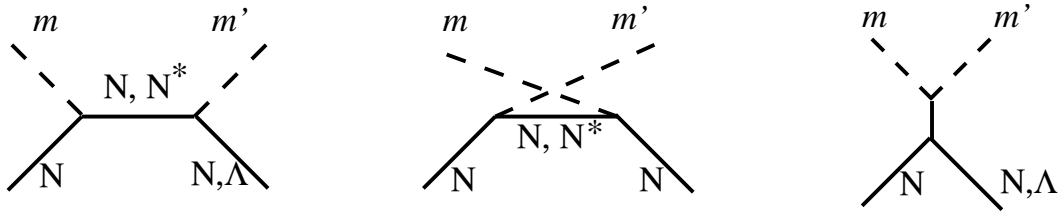}}
       \caption{
$s$-,$u$-, and $t$- channel contributions to the interaction potential. $m$ and $m'$ stands for
initial and final $\pi$, $\gamma$, $\omega$,  $K$, ...  etc.
      \label{diagr}}
  \end{center}
\end{figure}

The Bethe-Salpeter equation (BSE) is solved in the $K$-matrix
approximations to calculate the scattering amplitudes of different
reactions \cite{Penner:2002a,Penner:2002b,Penner:PhD}. The
validity of this approximation is discussed in
\cite{Penner:2002a,shklyar:2004b}. The interaction potential
($K$-matrix) in the BSE is constructed as a sum of the  $s$-, $u$-
and $t$-channel  tree-level Feynman diagrams contributions,
depicted in Fig.~\ref{diagr}, and calculated from the corresponding
effective interaction Lagrangians. After the partial wave
decomposition \cite{Penner:2002a,Penner:2002b,Penner:PhD} the BSE
reduces  to  the set of  algebraic  equations  for the scattering $T$-matrix:
\bea
T^{J\pm,I}_{fi} = \left[\frac{ K^{J\pm,I}}{1-iK^{J\pm,I}}\right]_{fi},
\label{GBS3}
\eea
where $J\pm$,$I$ are total spin, parity and izospin of the initial and final states
$f$,$i$=$\pi N$, $K \Lambda$, ... etc.

The resonance couplings to $K\Lambda$ are given  in
\cite{Penner:2002a,Penner:2002b,shklyar:2004a,shklyar:2004b}.
Each meson and baryon  vertex is
dressed by a corresponding form factor which according to \cite{Penner:2002a} is
chosen as:
\bea
F_p (q^2,m^2) &=& \frac{\Lambda^4}{\Lambda^4 +(q^2-m^2)^2},
\label{formfact}
\eea
thereby cutting off large 4-momenta $q^2\gg \Lambda^2$.

 To reduce the number of free parameters
we use the same cutoffs for all resonances with given spin $J$,
see \cite{shklyar:2004b}, e.g.,
$\Lambda^{N^*(1535)}_{i}$=$\Lambda^{N^*(1650)}_{j}$ where indices
$i$,$j$ run over all final states $i$,$j$=$\pi N$, $\eta N$,
$K\Lambda$, ...etc. Also, we identify the cutoff at the
$NK\Lambda$  vertex with the nucleon cutoff:
$\Lambda_{NK\Lambda}$=$\Lambda_{N}$=0.95 GeV.

The non-resonant part of the transition amplitude $(\pi,\gamma)N \to K\Lambda$
is similar to the one used in \cite{Penner:2002a,Penner:2002b,shklyar:2004b} 
and consists of the nucleon Born term and $t$-channel contributions 
with the $K^*$, $K^*_0$, and $K_1$ mesons
in the intermediate state. Taking the values for the decay widths from 
PDG \cite{pdg:2002},
the following couplings are extracted:
\bea
\ba{lcrclcr}
g_{K^* K \pi}          &=& -6.500 \; , & &
g_{K^*_0 K \pi}        &=& -0.900 \; ,  \\
g_{{K^*}^+ K^+ \gamma} &=& -0.414 \; , & &
g_{{K^*}^0 K^0 \gamma} &=&  0.631 \; , \\
g_{K_1^+ K^+ \gamma}   &=&  0.217 \; , & &
g_{K_1^0 K^0 \gamma}   &=&  0.217 \; . \\
\ea
\label{mesdeccons}
\eea
Note, that we use the same $\Lambda_t$=0.75 GeV at the corresponding $t$-channel 
vertices for both associated strangeness production and
non-strange channels \cite{shklyar:2004b}. Similar to our previous studies
\cite{Penner:2002a,Penner:2002b} we do  not include the $u$-channel diagrams
to the $(\pi,\gamma) N \to K\Lambda$ reaction.   The calculation of such contributions 
would require the knowledge of a priori unknown couplings to the intermediate 
strange baryons. To keep the model as
simple as possible,  these diagrams are not taken into account here.

\begin{table}[t]
  \begin{center}
    \begin{tabular}
      {l|l|l | l| l }
      $L_{2I,2S}$
& mass$^a$& $R_{K\Lambda}(C)$      & $R_{K\Lambda}(S)$     & $\bar R_{K\Lambda}$  \\
      \hline
      $S_{11}(1535)$
& 1526    &    $1.3^b$          &     $ 1.26^b$      &                          \\
      $S_{11}(1650)$
& 1664    &    $ 3.2(+)$        &     $ 4.6(+)$      &  4(1)             \\
      \hline
      $P_{11}(1440)$
& 1517    &    $ 1.48^b$        &     $-0.71^b$      &                          \\
      $P_{11}(1710)$
& 1723    &    $ 6.8( +)$       &     $ 3.1(+)$      &  5(3)              \\
      \hline
      $P_{13}(1720)$
& 1700    &    $ 4.6(+)$        &     $ 4.0(+)$      &  4.3 (0.4)                 \\
      $P_{13}(1900)$
& 1998    &    $ 2.4(+)$        &     $ 2.3(+)$      &  2.4 (0.3)                  \\
      \hline
      $D_{13}(1520)$
& 1505    &    $-0.58^b$        &     $-0.33^b$      &                           \\
      $D_{13}(1950)$
& 1934    &    $ 0.1(+)$        &     $ 0.1(-)$      &  0.1(0.1)                           \\
      \hline
      $D_{15}(1675)$
& 1666    &    $ 0.2(+)$        &     $ 0.1(+)$      &  0.1(0.1)                          \\
      \hline
      $F_{15}(1680)$
& 1676    &    $ 0.0(+)$        &     $ 0.0(+)$      &  0.1 (0.1)                         \\

      $F_{15}(2000)$
& 1946    &    $ 0.0(+)$        &     $ 0.2(-)$      &  0.1 (0.1)                        \\
    \end{tabular}
  \end{center}
  \caption{Branching decay ratios of  nucleon resonances into the  $K\Lambda$
    final state extracted in the calculations with $C$ and $S$ parameter sets, respectively.
    In brackets, the sign of corresponding coupling constant is shown (all $\pi N$
    couplings are chosen to be positive, see \cite{shklyar:2004b} ).
    In the last column
    the summary results for resonances with masses above the $K\Lambda$ threshold
    are given.  In brackets, the corresponding errors 
    are shown. The resonance mass is given in MeV, the
    decay ratios in percent.
    $^a$: fixed in the previous calculations \cite{shklyar:2004b}.
    $^b$: the coupling is given since the resonance mass is below the threshold.
    \label{tab12}}
\end{table}

\begin{table}[t]
  \begin{center}
    \begin{tabular}
      {l|r|l|r|l|r|l|r}
      \hhline{========}
      $g$ & value & $g$ & value & $g$ & value & $g$ & value \\
      \hhline{========}
$g_{N\Lambda K}$ & $-6.04$ & $g_{N\Lambda K_0^*}$ &  32.2  & $g_{N\Lambda K^*}$ & $ 2.28 $ & $\kappa_{N\Lambda K^*}$ &  $-0.01$ \\
                 & $-4.70$ &                      &  32.5  &                    & $ 7.00 $ &                         &  $-0.06$  \\
      \hline
      \hhline{========}
    \end{tabular}
  \end{center}
  \caption{Nucleon and $t$-channel couplings. First line: $C$-calculations.
Second  line: $S$-calculations (see text).
    \label{Born_coupl}}
\end{table}

\section{\label{details} Details of the calculations}
Our previous calculations on the associated strangeness production
\cite{shklyar:2004a,Penner:2002a,Penner:2002b} were based on the
experimental data published before 2002 (see  also
\cite{Penner:PhD} for details). Meanwhile, a new set of
photoproduction data has become available. This includes the
photon beam asymmetry obtained by SPring-8 \cite{Zegers:2003}, the
differential cross sections and the polarization of the outgoing
$\Lambda$  from SAPHIR \cite{Glander:2003} and CLAS
\cite{McNabb:2003}. Therefore, in the $\gamma N \to K\Lambda$
channel we do not use the previous SAPHIR data \cite{Tran:1998}
any longer but incorporate all recent measurements
\cite{Zegers:2003,Glander:2003,McNabb:2003} for energies
$\sqrt{s}\leq$ 2 GeV  into our database.

Comparing the $\gamma N \to K\Lambda$ differential cross  sections
independently  measured by SAPHIR \cite{Glander:2003} and  CLAS
\cite{McNabb:2003} one finds a significant disagreement between
the two data sets near 1.9 GeV. To avoid this problem, in a first
step only those data have been included into the fit which
coincide, within their error bars,  with each other. Using this
'truncated' data base, a full coupled-channel calculations on the
$\pi N \to \pi N$, $2\pi N$, $\eta N$, $\omega N$, $K\Lambda$,
$K\Sigma$ and $\gamma N \to \gamma N$, $\pi N$, $\eta N$, $\omega
N$, $K\Lambda$, $K\Sigma$ reactions  has been carried out. The
results of this calculations have been presented in
\cite{shklyar:2004b} focusing on the description of  the
non-strange channels. To pin down the $K\Lambda$ production
mechanism further, we constructed two different sets of
parameters. Set $S$ ($C$) corresponds to the solution obtained
with the differential $\gamma N \to K\Lambda$ cross section data
exclusively from the SAPHIR \cite{Glander:2003} (CLAS
\cite{McNabb:2003}) measurements.

At this step we allow all couplings to the $K\Lambda$ final state
to be varied during the fit. Other  parameters of the model which
correspond to the couplings to  the non-strange final states
($\gamma N$, $\pi N$ etc., see \cite{shklyar:2004b}) have been
held fixed at this stage.

Finally, we obtain two solutions  $S$  and $C$ which differ in 
their treatment of the $K\Lambda$ channel.
It will be seen later, that the main difference between these 
two solutions consists in the different
description of the background contributions to the $K\Lambda$ 
photoproduction. As a result, the additional
constraint from the SAPHIR or CLAS data hardly affects the non-strange 
channels. Thus, the  deviation from the $\chi^2$ obtained 
in \cite{shklyar:2004b} for the $\pi N$, $2\pi$, $\omega N$ 
etc does not exceed  $1.5\%$.

In the present study we obtain for the $K\Lambda$ photoproduction process
$\chi^{2}_{K\Lambda}$=2.0(2.2) in the $S$($C$)-calculation.
Note, that a comparison of  these values with the results of the previous calculations
\cite{Penner:2002b} should be taken with  care since the present study uses
different experimental input.

The couplings which have been varied in the fit are presented in Tables \ref{tab12}, \ref{Born_coupl}.
All other model parameters can be found in \cite{shklyar:2004b}.

\begin{figure}
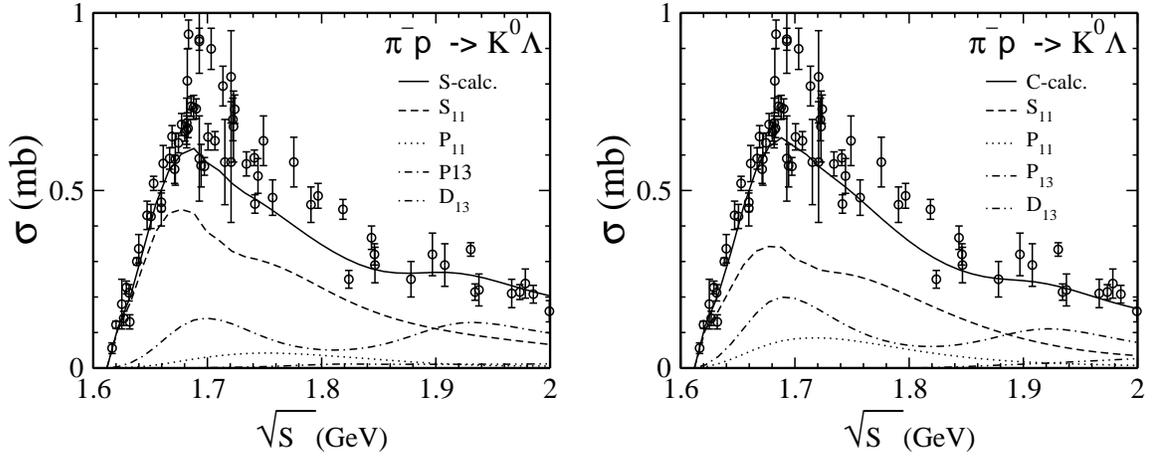

  \begin{center}
    \parbox{16cm}{
      \parbox{77mm}{\includegraphics*[width=77mm]{fig2.eps}}
      \parbox{77mm}{\includegraphics*[width=77mm]{fig3.eps}}
       }
       \caption{$\pi^- p\to K^0\Lambda$ 
total cross section.
Left(right) panel: partial wave cross sections  calculated using set $S$($C$).
Experimental data are taken from \cite{Baker:1978,Saxon:1979,Knasel:1975}.
      \label{fig2}}
  \end{center}
\end{figure}

\begin{figure}
  \begin{center}
{\includegraphics*[width=15cm]{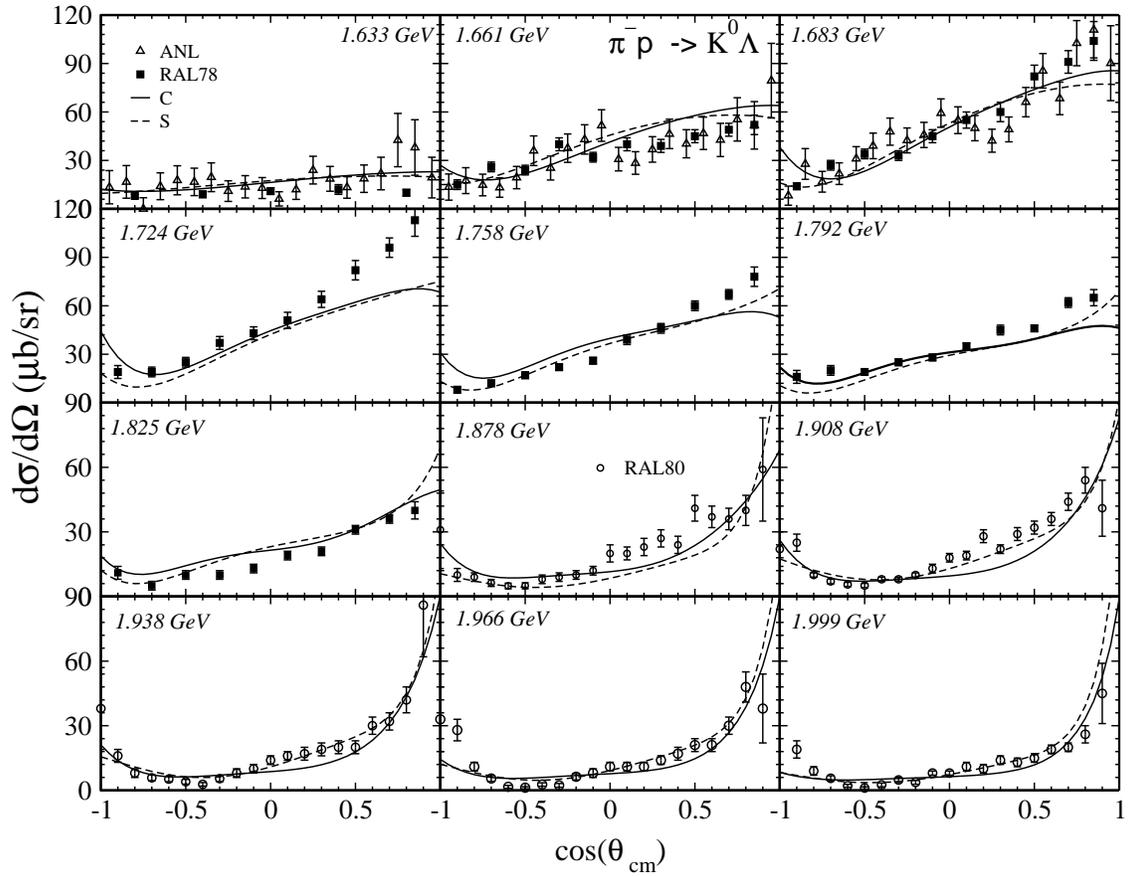}}
       \caption{
Comparison of the $\pi N \to K\Lambda$  differential cross section calculated using the $C$(solid line)
and $S$(dashed line) sets. Data are taken from
RAL78\cite{Baker:1978}, RAL80\cite{Saxon:1979}, and ANL\cite{Knasel:1975}.
      \label{fig3}}
  \end{center}
\end{figure}

\begin{figure}
  \begin{center}
{\includegraphics*[width=15cm]{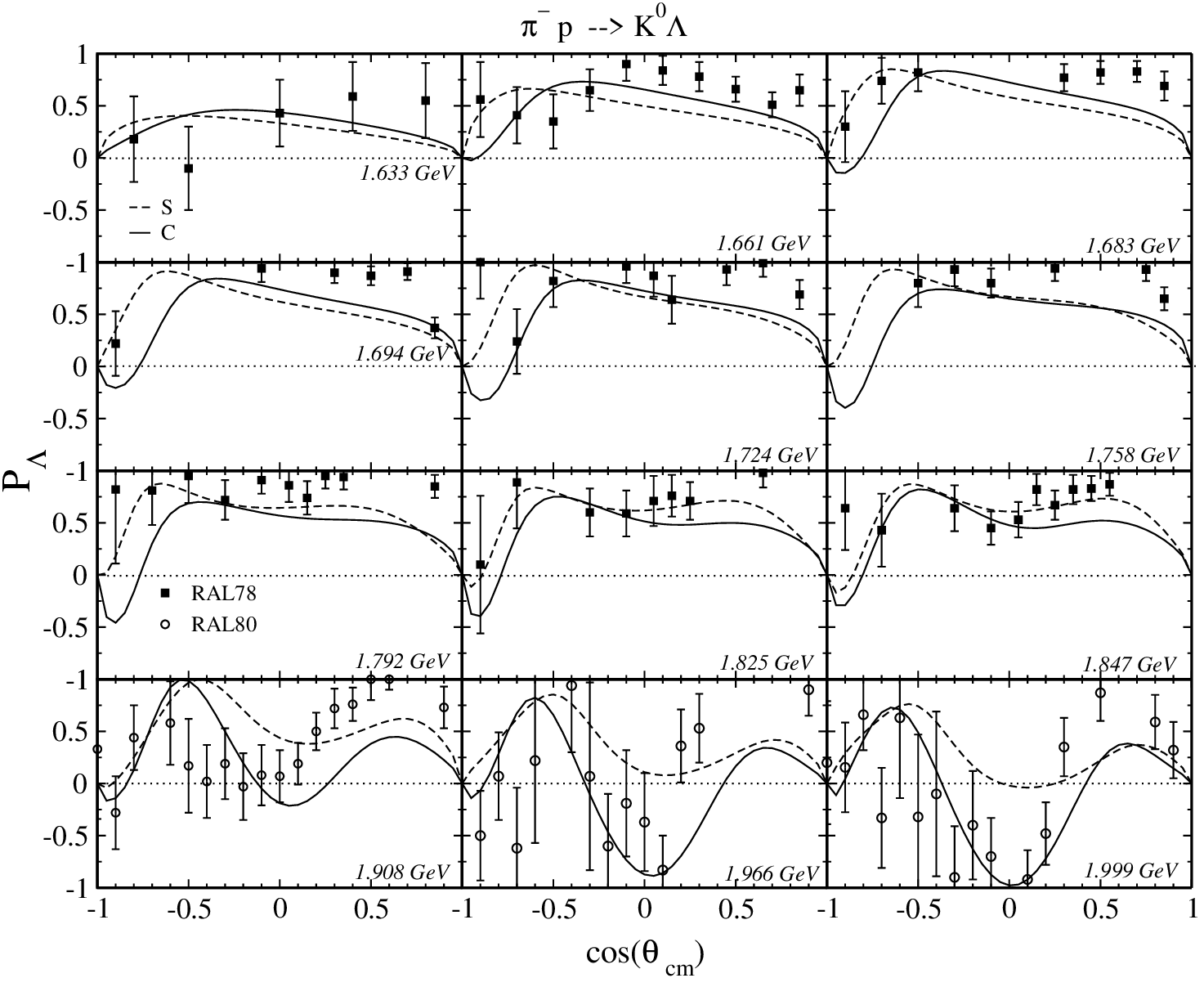}}
       \caption{
Comparison of $\Lambda$-polarization in the $\pi^- p\to K^0\Lambda$ reaction 
calculated for two different parameter sets.
Data are taken from  RAL78\cite{Baker:1978} and RAL80\cite{Saxon:1979}.
      \label{fig4}}
  \end{center}
\end{figure}

\section{\label{results}Results and discussion}
Since the recent  $K\Lambda$ photoproduction data
\cite{Glander:2003,McNabb:2003} give an indication for 'missing'
resonance  contributions, a combined analysis of the $(\pi,\gamma)
N \to K\Lambda$ reactions becomes inevitable to pin down these
states. Assuming small couplings to $\pi N$, these 'hidden' states
should not exhibit themselves in the pion-induced reactions and,
consequently, in the $\pi N \to K\Lambda $ reaction. The  aim of
the present calculations is to explore to what extent the new data
can be explained by known reaction mechanisms \cite{Penner:2002b},
without introducing new resonances.

The obtained nucleon resonance properties are presented in Table
\ref{tab12}. The decay ratios to the non-strange final states  and
the  electromagnetic  properties can be found in
\cite{shklyar:2004b}. We did not aim to distinguish between the
CLAS and SAPHIR data in the present calculations.  Performing
different calculations we only test the sensitivity of the
extracted resonance parameters to the various experimental input.

\subsection{\label{KL_hadr}$\pi N \to K\Lambda$}
We corroborate our previous findings \cite{Penner:2002a} where the major contributions to this
reaction are found to be from the $S_{11}$ and $P_{13}$ partial waves, see Fig. (\ref{fig2}).
However, opposite to \cite{Penner:2002a} the role of the $S_{11}$ partial wave becomes more
pronounced in the present calculations. Both $S$- and $C$-calculations give a similar description of
$\pi N \to K\Lambda$. The peaking behaviour observed in the  $S_{11}$ partial wave near
1.67 GeV, is induced by the  $S_{11}(1650)$ resonance.  The  $P_{13}$ wave consists of the  $P_{13}(1720)$
and $P_{13}(1900)$ resonance contributions which develop the two bumps  at 1.7 and 1.95 GeV, respectively.

Because of the mentioned  partial disagreement between the CLAS and SAPHIR photoproduction data,
the $S$- and $C$- calculations differ in their description of the non-resonance couplings to
$K\Lambda$. This leads to the different
background strengths to  the $S_{11}$, $P_{11}$, and $P_{13}$ partial waves leaving, however,
the  $P_{13}$(1720) and
$P_{13}(1900)$ resonance
couplings almost unchanged, see Table \ref{tab12}. Comparing  the $S$- and
$C$-parameter sets, the largest difference in the resonance parameters  is observed for the
$P_{11}(1710)$ state.  However, in the present calculations this  resonance
is found to be almost completely of inelastic origin with a small branching
ratio to ${\pi N}$ \cite{shklyar:2004b}.
Therefore, this state gives only a minor contribution to the reaction  and
the observed  difference in the $P_{11}$ partial wave  between $S$- and $C$-results is due to
the Born term and the $t$-channel exchange contributions.

Since the $S_{11}(1650)$ resonance dominates the reaction
mechanism near the threshold, the difference in  the non-resonance
part of the reaction also affects the properties of this state by
decreasing the relative decay width $R_{K\Lambda}(1650)$ to the
value $3.6$ in the $C$-calculations. We do not see any significant
effect from the $D_{13}(1985)$  state. This resonance  is included
in the present calculations, but its couplings to the $K\Lambda$
final state is found to be small, see Table~\ref{tab12}.

The calculated differential cross sections  corresponding to the  $S$- and $C$-coupling sets are
shown in Fig.\ref{fig3}. Both results show a good agreement with the experimental data in the whole
energy region. A difference between the two solutions is only found at forward
and backward scattering angles. This is due to the fact that the CLAS photoproduction cross sections
rise at backward angles which is not observed by the SAPHIR group (see discussion below).
At other scattering angles the $S$ and $C$ results are very similar.
The differences between $S$- and $C$-calculations are more pronounced 
for the $\Lambda$-polarization shown in
Fig. \ref{fig4}. Again, the main effect is seen at the backward angles where the polarization
changes its sign in the $C$-calculations.
Unfortunately, the quality of the data does not allow to pin down the reaction mechanism
further.

\subsection{\label{KL_photo}$\gamma N \to K\Lambda$}
The measurement of this reaction performed by Tran et. al.
\cite{Tran:1998} shows a resonance-like peak in the total
photoproduction cross section around 1.9 GeV. The new   data
published by the SAPHIR \cite{Glander:2003} and  CLAS
\cite{McNabb:2003} groups confirm the previous findings of
\cite{Tran:1998}. Moreover, due to the higher  resolution, the
peaking behaviour  in the differential cross sections at 1.9 GeV
has been found also for backward scattering angles. The
interpretation of this data is controversial in the literature.
The main question under discussion  is whether in  these
measurements contributions from presently unknown resonances are
observed or if they can be explained by already established
reaction mechanisms.

\begin{figure}
  \begin{center}
{\includegraphics*[width=17cm]{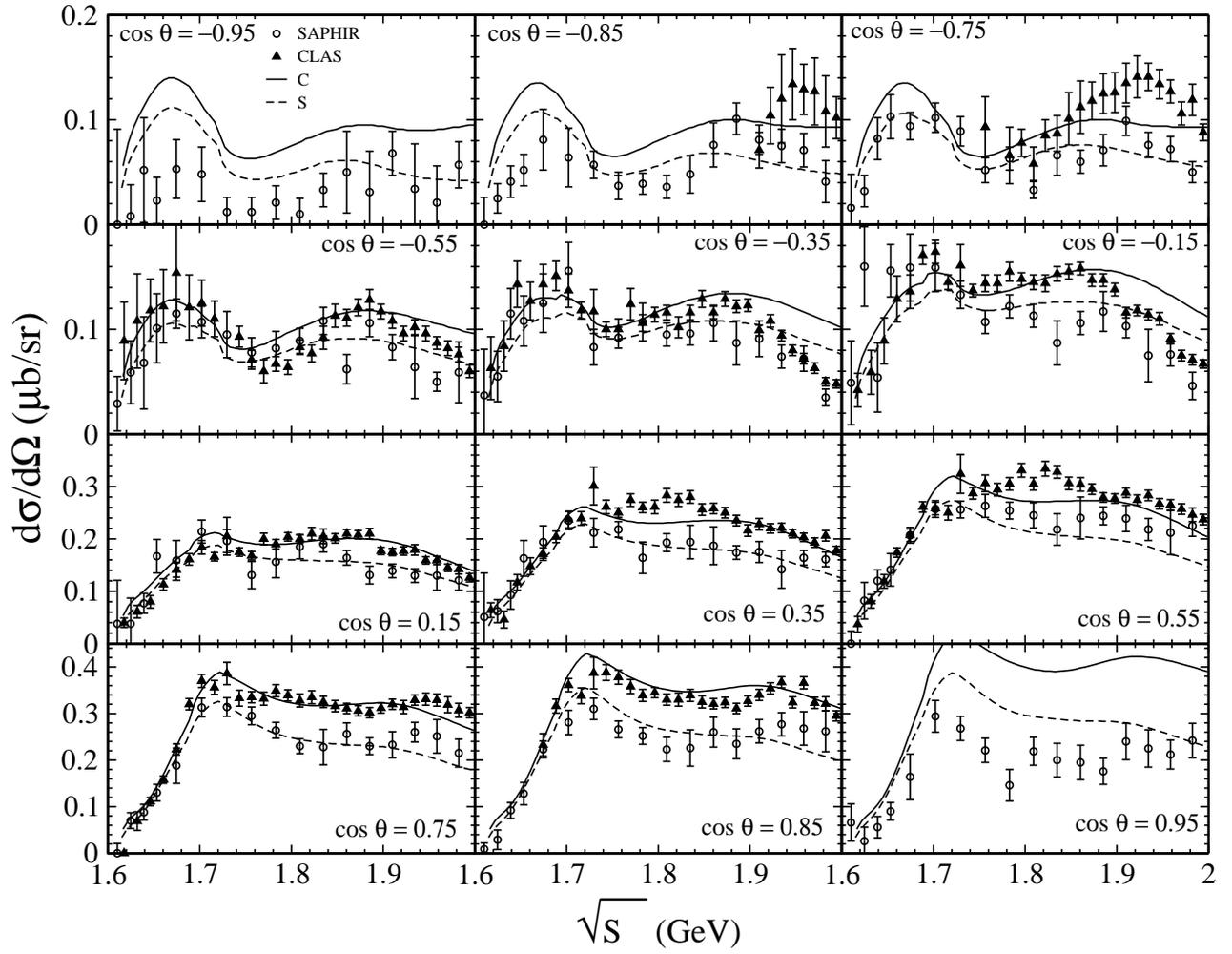}}
       \caption{
Comparison of the differential cross sections for the reaction
$\gamma p \to K^+\Lambda$ calculated using $C$ and $S$ parameter sets.
Experimental data are taken from \cite{McNabb:2003}(CLAS) and \cite{Glander:2003}(SAPHIR).
      \label{fig5}}
  \end{center}
\end{figure}

Since our previous investigations of the $K\Lambda$
photoproduction \cite{Penner:2002b} were based on the Tran et. al.
\cite{Tran:1998} data,  those  calculations lose the agreement
with the new data at 1.9 GeV for  the backward scattering angles.
Guided by the results of \cite{Penner:2002b} we have performed a
new coupled-channel study of this reaction using separately the
CLAS and SAPHIR measurements as two independent input sets, see
Section \ref{details}. The main  difference between the CLAS and
SAPHIR data is seen at backward and forward directions,
Fig.~\ref{fig5}. Both measurements show two peaks but disagree in
the absolute values of the corresponding differential cross
sections. Also, the second bump in the CLAS data is shifted to the
lower energy 1.8 GeV for the scattering angles corresponding to
$\cos\theta$=0.35 and $\cos\theta$=0.55.

Both the $S$ and $C$ calculations demonstrate a good agreement
with the corresponding experimental data, although the
$S$-calculations lead to a smaller value of $\chi^2_{K\Lambda}$.
However, this is not because the measurements of
\cite{Glander:2003} are more 'consistent' with  $\pi N \to
K\Lambda$ data.  Instead it is due to the fact that the location of
the second peak in the CLAS data changes with scattering angles.
Since our differential cross section does not follow the CLAS data
at 1.8 GeV and $\cos\theta$=0.35 and $\cos\theta$=0.55 (see
Fig.~\ref{fig5})  the total $\chi^2_{K\Lambda}$ turns out to be
larger in the $C$-calculations. Note, that if the behaviour
observed by the CLAS group should be confirmed in future
experiments, further assumptions on the reaction mechanism would
be required.

\begin{figure}
  \begin{center}
{\includegraphics*[width=14cm]{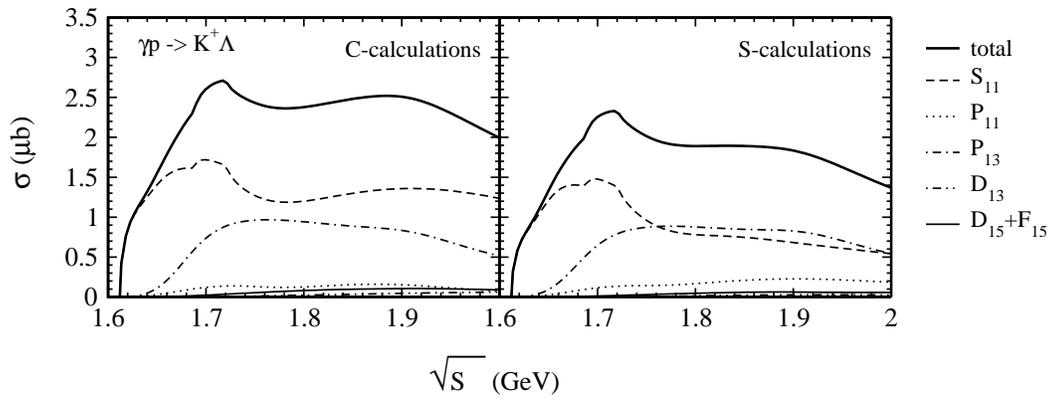}}
       \caption{
Partial wave contributions to the total $K\Lambda$ photoproduction cross section.
      \label{fig6}}
  \end{center}
\end{figure}

\begin{figure}
  \begin{center}
{\includegraphics*[width=14cm]{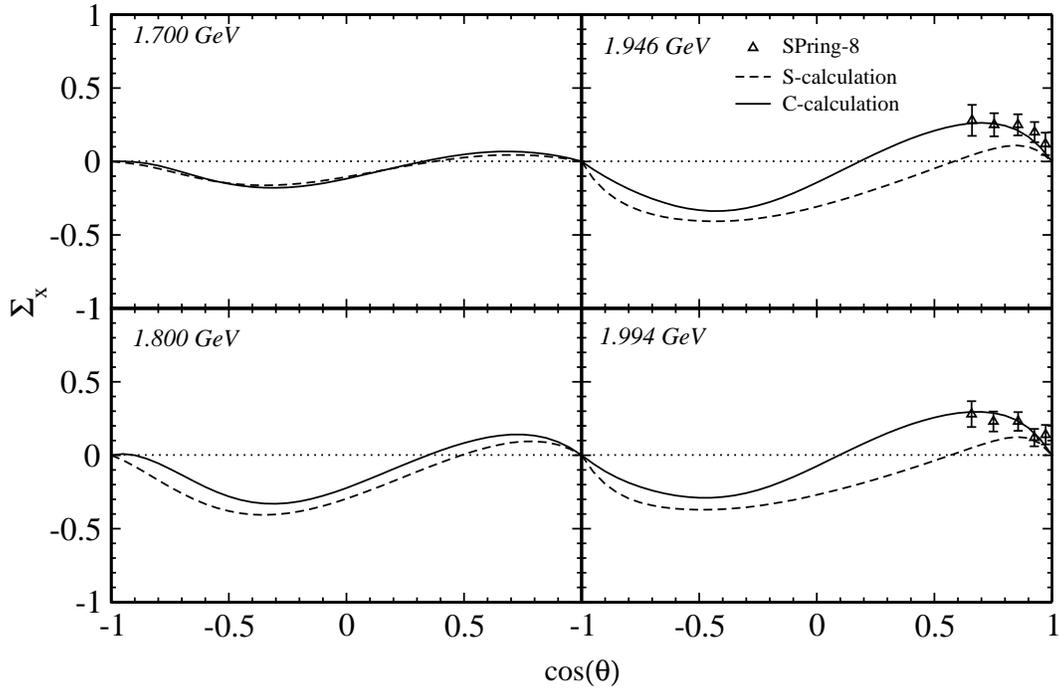}}
       \caption{
The calculated photon beam asymmetry. Data are taken from \cite{Zegers:2003}
      \label{fig7}}
  \end{center}
\end{figure}

\begin{figure}
  \begin{center}
{\includegraphics*[width=14cm]{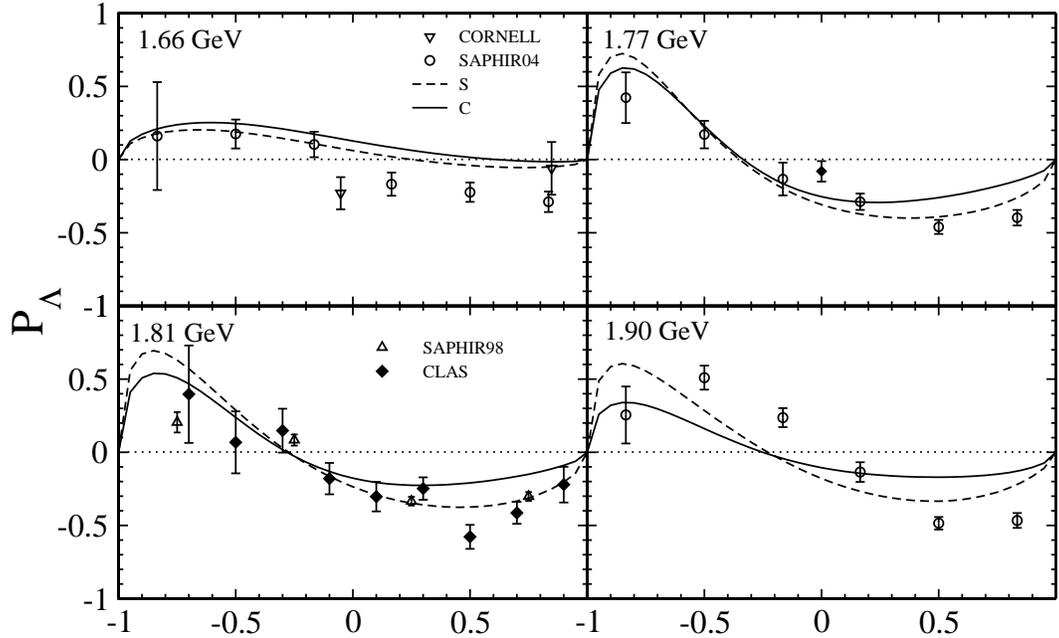}}
       \caption{
$\Lambda$-polarization in the $\gamma p \to K^+\Lambda$ reaction. 
Data are from  SAPHIR98\cite{Tran:1998},
SAPHIR04\cite{Glander:2003}, CLAS \cite{McNabb:2003}, 
CORNELL\cite{Cornell:1963}.
      \label{fig8}}
  \end{center}
\end{figure}

Similar to  $\pi N \to K\Lambda$ the major difference between the
$S$ and $C$ solutions is the treatment of the non-resonant
contributions. Thus, both calculations show  two peaks in the
differential cross sections at 1.7 and 1.9 GeV. The first  bump at
1.67 GeV in both calculations is produced by the $S_{11}(1650)$
resonance, see Fig.~\ref{fig6}. The relative contributions to the
second peak  at 1.9 GeV are different in the $C$ and $S$
solutions. In the $C$-calculations this structure is described by
the $S_{11}$ partial wave. Note, that we do not include a third
$S_{11}$ resonance  with a mass of about 2 GeV as done in
\cite{JuliaDiaz:2005}. Therefore, the contributions to the
$S_{11}$ at higher energies are dominated by the non-resonant
reaction  mechanisms.  The $P_{13}$ partial wave is entirely
driven by the $P_{13}(1720)$ and $P_{13}(1900)$ resonance
contributions. Switching off these resonance couplings to
$K\Lambda$  leads to an almost vanishing  $P_{13}$ partial wave.
In the $S$-calculations no peaking behaviour is found in the
$S_{11}$ partial wave at 1.95 GeV. However, the non-resonant
effects in the $S_{11}$ channel are still important. The role of
the $P_{13}$ resonances are slightly enhanced  in the
$S$-calculations. The effect from the $P_{11}(1710)$ resonance is
found to be small  in both calculations  due to  destructive
interference with the background process. There are no significant
contributions from the spin-$\ffh$ resonances to the $\gamma N \to
K \Lambda$ reaction. Also, no any effect is seen from the
$D_{13}(1950)$ resonance.

The calculated photon beam asymmetry  $\Sigma_x$  and  recoil
polarization $P_\Lambda$ are shown in Fig.~\ref{fig7} and Fig.~\ref{fig8}.
Since the beam asymmetry  data \cite{Zegers:2003} from the
SPring-8 collaboration are available only for energies above 1.94
GeV, these measurements give  an insignificant  constraint on the
model parameters. Therefore, the results for  the asymmetry might
be regarded as a prediction rather than an outcome of the fit.
More information comes from the $\Lambda$-polarization data. A
good description of the $\Sigma_x$ and $P_\Lambda$ data is
possible in both the $C$ and $S$ calculations. One can conclude,
that despite the differences in the differential cross sections
between the data \cite{Glander:2003,McNabb:2003} the calculated
$\Sigma_x$ and $P_\Lambda$ observables are very similar in both
cases and the main difference between the $S$- and
$C$-calculations consists in the background contributions.
This also explains why  the resonance parameters extracted 
are fairly insensitive to the parameter set used (see Table \ref{tab12}).

\section{\label{summary} Summary  }
In summary, we have  performed a coupled-channel analysis of the
$(\pi, \gamma) N \to K\Lambda$ reaction to extract
the non-strange resonance couplings to the $K\Lambda$ final state.
To distinguish between the different $K\Lambda$ photoproduction 
measurements we obtained two independent solutions to the SAPHIR 
and CLAS data for energies $\sqrt{s}\leq$ 2 GeV.
The main resonance contributions to the reaction stem from the 
$S_{11}(1650)$, $P_{13}(1720)$, and $P_{13}(1900)$ states.  
It is shown that  the extracted resonance parameters are hardly 
sensitive to the observed discrepancy between the different  data 
sets. We have discussed that this is due to the fact that the
differences between the two data sets stem mainly from different 
non-resonant background contributions.

We do not see any significant effects from the  $P_{11}(1710)$ and
$D_{13}(1890)$ states. Also, the contributions from the
spin-$\ffh$ resonances are found to be small. In our
coupled-channel approach the second bump in the differential cross
section data  at 1.9 GeV observed by the SAPHIR and CLAS groups is
produced by a coherent sum of the resonance and background
contributions, without any evidence for a 'missing' resonance.
As a test for our model calculations we predict the beam asymmetry 
to change its sign for the moderate angles. This effect can be easily 
checked at the running experimental facilities such as
JLAB and LEPS.

We have  checked whether there is room left for a further
improvement of the agreement of the calculated observables with
the $(\pi,\gamma) N \to K\Lambda$  data. 
However, before such a new
analysis is meaningful the inconsistency between the two data sets
has to be resolved.

\acknowledgments

 The work has been supported by Forschungszentrum Juelich.

\bibliographystyle{h-physrev3}
\bibliography{tau}

\begin{thebibliography}{10}

\bibitem{Capstick:1998uh}
S.~Capstick and W.~Roberts,
\newblock Phys. Rev. {\bf D58}, 074011 (1998), nucl-th/9804070.

\bibitem{Tran:1998}
SAPHIR, M.~Q. Tran {\em et~al.},
\newblock Phys. Lett. {\bf B445}, 20 (1998).

\bibitem{Penner:2002b}
G.~Penner and U.~Mosel,
\newblock Phys. Rev. {\bf C66}, 055212 (2002), nucl-th/0207069.

\bibitem{Mart:1999}
T.~Mart and C.~Bennhold,
\newblock Phys. Rev. {\bf C61}, 012201 (2000), nucl-th/9906096.

\bibitem{JuliaDiaz:2005}
B.~Julia-Diaz {\em et~al.},
\newblock (2005), nucl-th/0501005.

\bibitem{McNabb:2003}
The CLAS, J.~W.~C. McNabb {\em et~al.},
\newblock Phys. Rev. {\bf C69}, 042201 (2004), nucl-ex/0305028.

\bibitem{Mart:2004}
T.~Mart, A.~Sulaksono, and C.~Bennhold,
\newblock Talk at International Symposium on Electrophoto Production of
  Strangeness on Nucleons and Nuclei (SENDAI 03), Sendai, Japan, 16-18 Jun
  2003.  (2004), nucl-th/0411035.

\bibitem{Janssen:2001}
S.~Janssen, J.~Ryckebusch, D.~Debruyne, and T.~Van~Cauteren,
\newblock Phys. Rev. {\bf C65}, 015201 (2002), nucl-th/0107028.

\bibitem{Ireland:2004}
D.~G. Ireland, S.~Janssen, and J.~Ryckebusch,
\newblock Nucl. Phys. {\bf A740}, 147 (2004).

\bibitem{shklyar:2004a}
V.~Shklyar, G.~Penner, and U.~Mosel,
\newblock Eur. Phys. J. {\bf A21}, 445 (2004), nucl-th/0403064.

\bibitem{Penner:2002a}
G.~Penner and U.~Mosel,
\newblock Phys. Rev. {\bf C66}, 055211 (2002), nucl-th/0207066.

\bibitem{Chiang:2001}
W.-T. Chiang, F.~Tabakin, T.~S.~H. Lee, and B.~Saghai,
\newblock Phys. Lett. {\bf B517}, 101 (2001), nucl-th/0104052.

\bibitem{Lutz:2001}
M.~F.~M. Lutz, G.~Wolf, and B.~Friman,
\newblock Nucl. Phys. {\bf A706}, 431 (2002), nucl-th/0112052.

\bibitem{Usov:2005}
A.~Usov and O.~Scholten,
\newblock (2005), nucl-th/0503013.

\bibitem{Saghai:2001}
B.~Saghai,
\newblock (2001), nucl-th/0105001.

\bibitem{Penner:PhD}
G.~Penner,
\newblock PhD thesis (in English), Giessen, 2002, available via
  http://theorie.physik.uni-giessen.de .

\bibitem{shklyar:2004b}
V.~Shklyar, H.~Lenske, U.~Mosel, and G.~Penner,
\newblock Phys. Rev. C. (in print) , nucl-th/0412029.

\bibitem{Zegers:2003}
LEPS, R.~G.~T. Zegers {\em et~al.},
\newblock Phys. Rev. Lett. {\bf 91}, 092001 (2003), nucl-ex/0302005.

\bibitem{Glander:2003}
K.~H. Glander {\em et~al.},
\newblock Eur. Phys. J. {\bf A19}, 251 (2004), nucl-ex/0308025.

\bibitem{Janssen:2001pe}
S.~Janssen, J.~Ryckebusch, W.~Van~Nespen, D.~Debruyne, and T.~Van~Cauteren,
\newblock Eur. Phys. J. {\bf A11}, 105 (2001), nucl-th/0105008.

\bibitem{Janssen:2003zv}
S.~Janssen, D.~G. Ireland, and J.~Ryckebusch,
\newblock Phys. Lett. {\bf B562}, 51 (2003), nucl-th/0302047.

\bibitem{Feuster:1998a}
T.~Feuster and U.~Mosel,
\newblock Phys. Rev. {\bf C58}, 457 (1998), nucl-th/9708051.

\bibitem{Feuster:1998b}
T.~Feuster and U.~Mosel,
\newblock Phys. Rev. {\bf C59}, 460 (1999), nucl-th/9803057.

\bibitem{Chiang:2004}
W.-T. Chiang, B.~Saghai, F.~Tabakin, and T.~S.~H. Lee,
\newblock Phys. Rev. {\bf C69}, 065208 (2004), nucl-th/0404062.

\bibitem{Arndt:2003}
R.~A. Arndt, W.~J. Briscoe, I.~I. Strakovsky, R.~L. Workman, and M.~M. Pavan,
\newblock Phys. Rev. {\bf C69}, 035213 (2004), nucl-th/0311089.

\bibitem{pdg:2002}
Particle Data Group, K.~Hagiwara {\em et~al.},
\newblock Phys. Rev. {\bf D66}, 010001 (2002),
\newblock \url{http://pdg.lbl.gov}.

\bibitem{Baker:1978}
R.~D. Baker {\em et~al.},
\newblock Nucl. Phys. {\bf B141}, 29 (1978).

\bibitem{Saxon:1979}
D.~H. Saxon {\em et~al.},
\newblock Nucl. Phys. {\bf B162}, 522 (1980).

\bibitem{Knasel:1975}
T.~M. Knasel {\em et~al.},
\newblock Phys. Rev. {\bf D11}, 1 (1975).

\bibitem{Cornell:1963}
H.~Thom, E.~Gabathuler, E.~Jones, B.~McDaniel, and W.~M. Woodward,
\newblock Phys. Rev. Lett. {\bf 11}, 433 (1963).

\end{thebibliography}

\end{document}